%% file: 0.root.tex
\documentclass[conference]{IEEEtran}
\IEEEoverridecommandlockouts
\usepackage[table,xcdraw]{xcolor}
\usepackage{url}
\usepackage{cite}
\usepackage{amsmath,amssymb,amsfonts}
\usepackage{algorithmic}
\usepackage{graphicx}
\usepackage{textcomp}
\usepackage{xcolor}
\def\BibTeX{{\rm B\kern-.05em{\sc i\kern-.025em b}\kern-.08em
    T\kern-.1667em\lower.7ex\hbox{E}\kern-.125emX}}

\usepackage[utf8x]{inputenc}
\usepackage{graphicx}
\usepackage{float}

\usepackage{fancyhdr}
\begin{document}

\title{Federated deep transfer learning for EEG decoding using multiple BCI tasks  \\
\thanks{Preprint accepted by 2023 IEEE NER [\textcopyright 2023 IEEE. Personal use of this material is permitted. Permission from IEEE must be obtained for all other uses, in any current or future
media, including reprinting/republishing this material for advertising or promotional purposes, creating new collective works, for resale or redistribution to servers or lists, or reuse of any copyrighted component of this work in other works.] Address for correspondence: aldo.faisal@imperial.ac.uk We acknowledge funding from UKRI Turing AI Fellowship to AAF (EP/V025449/1).}}

\author{\IEEEauthorblockN{Xiaoxi Wei}
\IEEEauthorblockA{\textit{Dept. of Computing, Imperial College London} \\
\textit{Brain \& Behaviour Lab}\\
London, UK\\
xiaoxi.wei18@imperial.ac.uk}
\and
\IEEEauthorblockN{A. Aldo Faisal}
\IEEEauthorblockA{
\textit{Brain \& Behaviour Lab, Imperial College London}\\
London, UK\\
% \textit{Institute of Artificial \& Human Intelligence, University of Bayreuth} \\
\textit{Chair in Digital Health \& Data Science, University of Bayreuth} \\
Bayreuth, Germany\\
aldo.faisal@imperial.ac.uk}
}

\maketitle

\begin{abstract}
Deep learning is the state-of-the-art in BCI decoding. However, it is very data-hungry and training decoders requires pooling data from multiple sources. EEG data from various sources decrease the decoding performance due to negative transfer~\cite{wei2021inter}. Recently, transfer learning for EEG decoding has been suggested as a remedy \cite{jayaram2016transfer,lotte2018review} and become subject to recent BCI competitions (e.g. BEETL~\cite{pmlr-v176-wei22a}), but there are two complications in combining data from many subjects. First, privacy is not protected as highly personal brain data needs to be shared (and copied across increasingly tight information governance boundaries). Moreover, BCI data are collected from different sources and are often with different BCI tasks, which has been thought to limit their reusability. Here, we demonstrate a federated deep transfer learning technique, the Multi-dataset Federated Separate-Common-Separate Network (MF-SCSN) based on our previous work of SCSN~\cite{wei2021inter}, which integrates privacy-preserving properties into deep transfer learning to utilise data sets with different tasks. This framework trains a BCI decoder using different source data sets from different imagery tasks (e.g. some data sets with hands and feet, vs others with single hands and tongue, etc). Therefore, by introducing privacy-preserving transfer learning techniques, we unlock the reusability and scalability of existing BCI data sets. We evaluated our federated transfer learning method on the NeurIPS 2021 BEETL competition BCI task. The proposed architecture outperformed the baseline decoder by 3\%. Moreover, compared with the baseline and other transfer learning algorithms, our method protects the privacy of the brain data from different data centres.
\end{abstract}

\input{1introduction.tex}
\input{2methods.tex}

\input{3evaluation.tex}

\input{4results.tex}
\input{5discussion.tex}

\bibliographystyle{IEEEtran}
\bibliography{reference}

\end{document}

%% file: 1introduction.tex
\section{Introduction}
Deep Learning based EEG decoding has become a standard in BCI \cite{walker2015deep, BallCNN} and unlocked state-of-the-art machine learning ideas to benefit neural engineering research. Deep learning approaches are usually data-hungry. Some previous studies deal with the lack of available EEG data by data-efficient approaches \cite{ferrante2015data,ortega2018compact,ponferrada2018data}. In recent years, the development of transfer learning in EEG decoding~\cite{jayaram2016transfer,lotte2018review} enables algorithms to learn more from combining different EEG data sets. However, most methods focus on only a single data set with a unified experiment setup or task. The scale of EEG data sets is usually limited to dozens of subjects, unlike biomedical data sets with thousands, due to the difficulty and cost of EEG data collection. The international BEETL EEG competition~\cite{pmlr-v176-wei22a} held at NeurIPS 2021 focused on cross-dataset EEG transfer learning and brought academic attention to utilising many EEG data sets across tasks for transfer learning with around 30 international competing teams. Several successful algorithms were proposed to tackle the BEETL challenge, which has provided fundamental design principles for cross-dataset and cross-task EEG transfer learning. With examples showing that heterogeneous EEG data sets from different data centres and sources could be utilised for large-scale machine learning algorithms, EEG data sharing privacy becomes the next concern. Brainwaves contain rich privacy information that could be potentially decoded, e.g. words and identities. Moreover, health data sharing across sites or countries is under strict legal governance restrictions \cite{sullivan2019eu}.

There were some strategies for privacy-preserving in the machine learning literature. Federated learning~\cite{li2020review} trains models on edge servers without exchanging the data. Data encryption methods encrypt raw data or parameters of the model~\cite{hao2019towards}. This requires an encoding-decoding procedure which introduces extra computational cost. Users also need to decrypt the data following a certain protocol given by the algorithm provider, in which case the protocol could be potentially hacked. Similarly, transformation~\cite{lyu2017privacy} adds cancelable noise to local gradients or parameters before uploading them to a central server. Methods of model splitting~\cite{dong2017dropping} are based on allocating different parameters to different data sets, thus protecting the model privacy of each individual. Multi-party computation \cite{du2001secure} trains models locally first, and then aggregation is done securely by a third party. However, all these methods are either unproven or unsuitable for EEG decoding, e.g. adding cancellable noise is problematic to the low signal-to-noise ratio of EEG. Moreover, they add computational burden, e.g. due to encryption, which further disadvantages  training on large-scale EEG data.

\begin{figure*}[htbp]
\centering
\vspace{ -5mm }
\includegraphics[width=1.7\columnwidth]{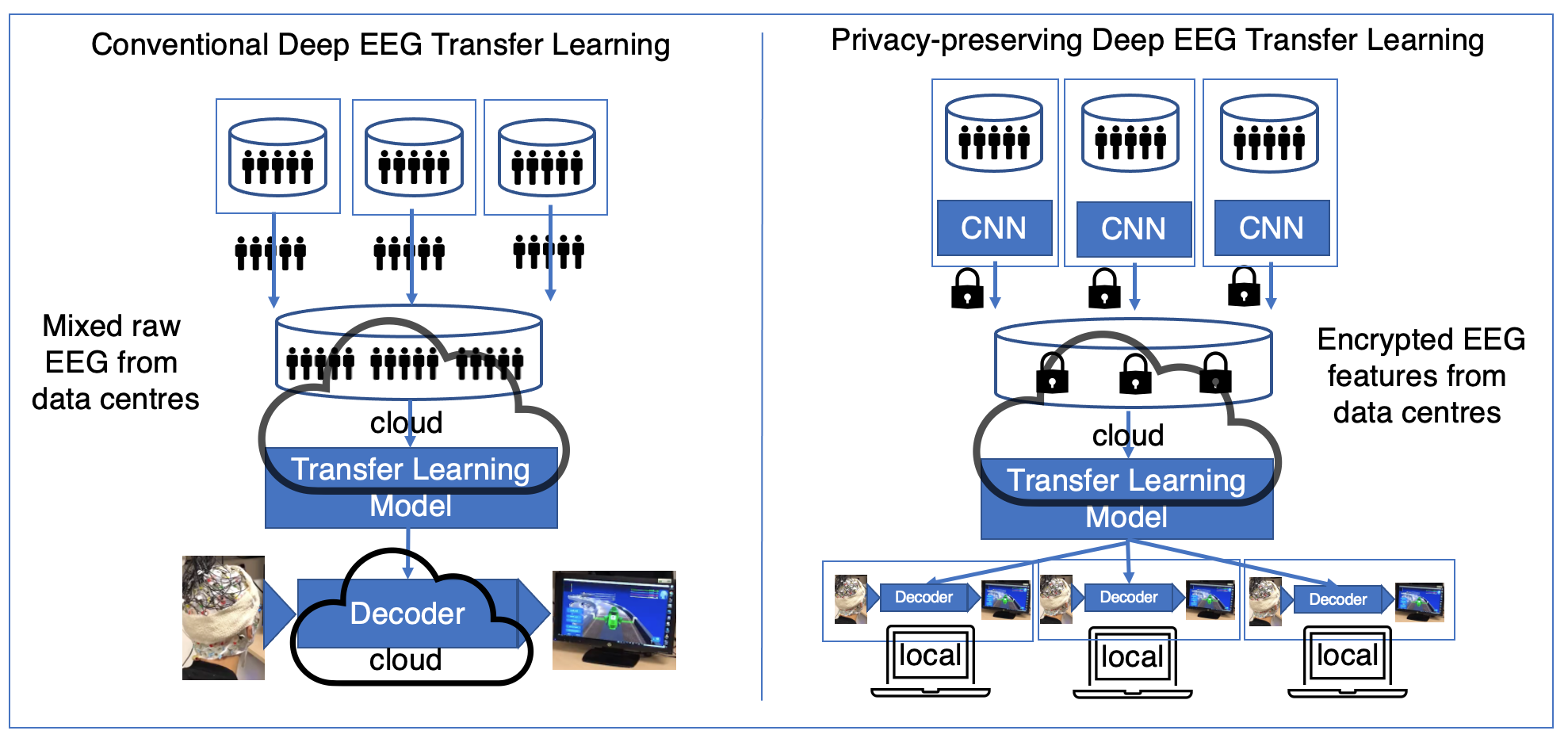}
\vspace{-3mm}
\caption{The figure illustrates the differences between conventional deep transfer learning and privacy-preserving deep transfer learning with federated models.}
\label{fig1}
\vspace{-6mm}
\end{figure*}

Privacy-preserving is drawing increasing attention in EEG decoding with the development of more accurate human intention decoders. There are a few studies in the EEG literature on privacy-preserving based on the above methods~\cite{popescu2021privacy,xia2022privacy,ju2020federated,bethge2022domain}, or based on better protocol or user level system design~\cite{kapitonova2022framework}. Our previous work and some other recent studies~\cite{wei2021inter,bethge2022domain} utilised distributed feature extractors to deal with individual EEG differences while maintaining private information without extra cost for encryption. However, both studies can not handle learning from different tasks and protect inference-level privacy, thus limiting the use of large-scale data. Therefore, it is still a challenge for cross-data-centre transfer learning with different tasks in a privacy-preserving way.

In this study, we propose an architecture to combine privacy-preserving machine learning with deep transfer learning. The Multi-dataset Federated Separate-Common-Separate Network (MF-SCSN) integrates privacy-preserving properties into deep transfer learning EEG decoding on multiple tasks.

%% file: 2methods.tex
\section{Method Development}
A benchmark architecture, the shallow ConvNet~\cite{BallCNN}, is used as the baseline model. The network includes a temporal layer to extract time-scale information, followed by a spatial layer to extract cross-channel features. Square non-linearity, average pooling and log non-linearity are then performed.

In this study, a privacy-preserving cross-dataset deep transfer learning architecture, the MF-SCSN, is proposed based on our previous study on inter-subject deep transfer learning~\cite{wei2021inter}. The variability of EEG comes from several aspects. There are superficial variabilities like sensor locations, sensor impedance and devices. These differences could be handled in shallow layers of a transfer learning network. For the intrinsic variability of individual brains, functionalities could be potentially learnt and handled more precisely in deeper layers. 

\begin{figure}
 \centering
  % \vspace{ -2mm }
 \includegraphics[width=1\columnwidth]{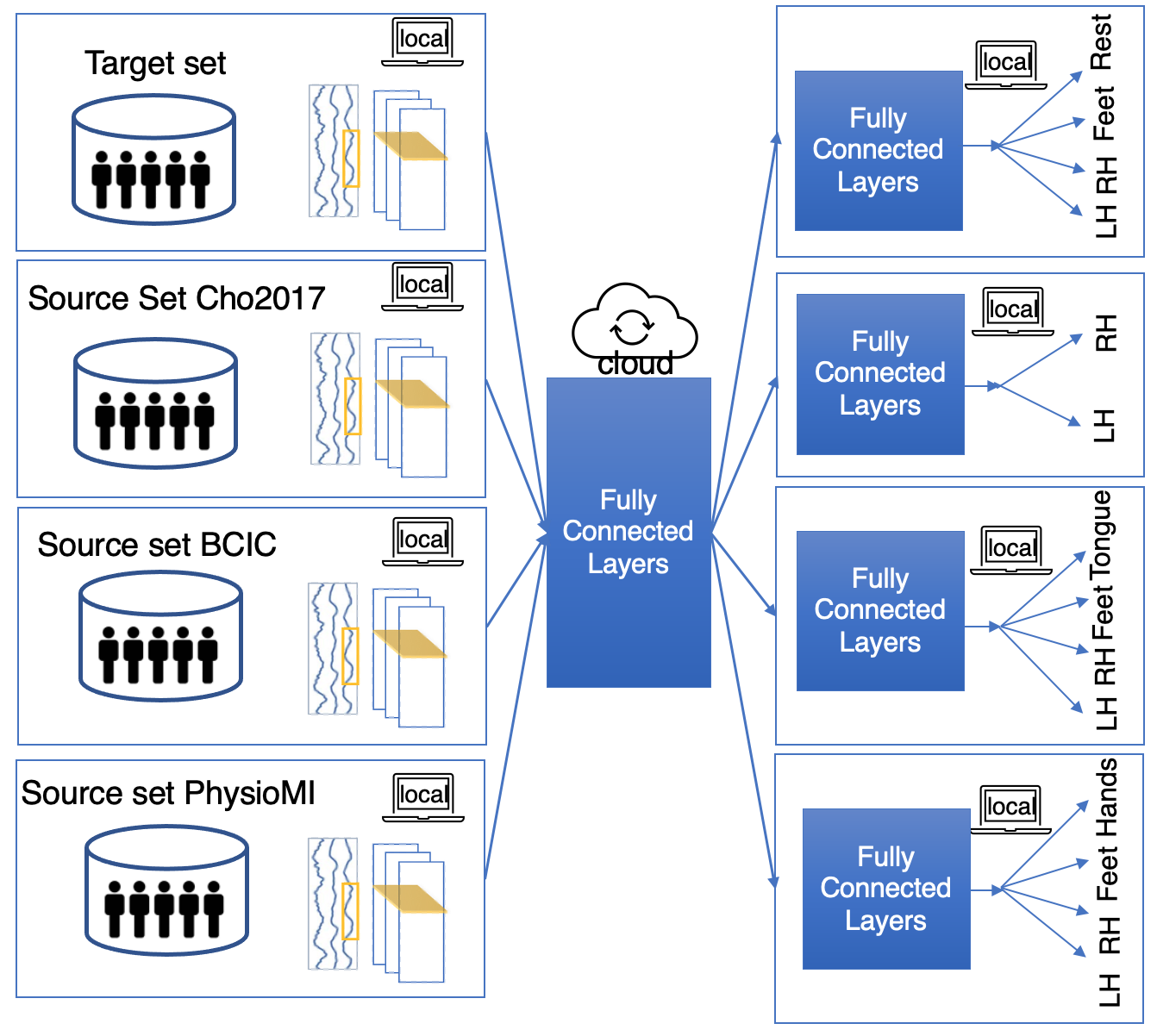} 
 \vspace{-8mm}
 \caption{Multi-dataset Federated Separate-Common-Separate Network (MF-SCSN). Raw data, feature extractors (the encryption keys), labels and predictions are all preserved locally.}
 \label{fig2}
 \vspace{-6mm}
\end{figure}

The MF-SCSN architecture consists of three main components (see Fig.~\ref{fig2}) it separates into both shallow layers and deeper layers to handle the variabilities while performing a joint feature extractor to learn common transferable knowledge across data sets.
The first set of components is the local branches (left side of the figure) as both feature extractors and `keys' for data encryption. The shallow ConvNet above is used here as a feature extractor. Raw data from different data centres is encrypted into EEG features through the local branches. Local servers of data centres conduct the computation of feature extraction, and parameters (keys) are stored locally. In this way, the proposed architecture preserves both data-level privacy and parameter-level privacy.

The second component of the MF-SCSN is a common transfer network located in a cloud server. This is where transfer learning across different data sets happens. Encrypted features are received from data centres in the cloud for common feature extraction. Previous local feature extractors handle variabilities in data set distributions. The design purpose of common layers is to find a common distribution to which different data sets and distributions could transfer to. Unlike other encryption methods for privacy-preserving, the MF-SCSN has no encryption costs since the federated feature extractors encrypt data automatically. Moreover, it does not have a `decryption' procedure because the encrypted features are exactly the inputs required by the common transfer layer. Therefore, the cloud server does not need to know any information about the local feature extractors (the `keys'). This further increases the parameter-level security of the model.

Finally, the transferred features are delivered to the third component of MF-SCSN, i.e. the deep separate layers and heterogeneous classifiers. Its design is motivated by \cite{pmlr-v176-wei22a} showing that combining different classifiers for cross-dataset transfer learning can help overcome label inconsistency. In light of this, the third set of components contains some separate layers to deal with further differences across data sets, followed by local classifiers specified for different tasks. Besides the benefit of handling label inconsistency, predictions and labels are preserved locally in this way. 

% Therefore, the Multi-dataset Federated SCSN preserves data-level, parameter-level and inference-level privacy as well as integrates fundamental deep transfer learning principles into the design.

%% file: 3evaluation.tex
\section{Evaluation Method}
The proposed architecture was tested on the motor imagery task of the BEETL competition. Details of the data can be found in~\cite{pmlr-v176-wei22a}. We used three source data sets with very different devices and data collection protocols. 
% \aldo{i see only 2 data sets here}
% \xiaoxi{it's BCICIV2a, It's not split into two paragraphs so was hidden.}
In general, we curated the data, by choosing subsets of subjects from each data set that had higher data quality, narrowly defined as embedding better discriminative features. Therefore, data quality per subject was determined from classification performance (independent of our work), as they were reported by previous studies or our replications of those studies.
BCICIV2a~\cite{BCIC} contains nine subjects performing left-hand, right-hand, feet and tongue motor imagery. Subject 1,3,7,8 and 9 were selected as sources. 
Cho2017~\cite{Cho2017} has 52 subjects performing left and right-hand motor imagery, all subjects but 32,46 and 49 were used.
PhysioMI~\cite{schalk2004bci2000,11} had 109 subjects, we selected as sources subject 1,7,17,24,28,31,33,34,35,42,49,52,
54,55,56,60,62,63,68,71,72,73,85,91,93,94,103. 
Test sets consist of 5 subjects with 200 trials each (1000 trials in total). Labels are left-hand, right-hand and feet motor imagery and rest state (250 trials each). 

All final accuracies are reported on the weighted accuracy of the three classes - left-hand (LH), right-hand (RH) and `other'. As described, source data sets and the target set have different classification tasks. Some data sets shared left/right hand or feet motor imagery as the target data sets, others have tongue and rest. Data with different labels have various distributions, which introduces extra complexity for transfer learning.

To align and make use of all data sets, we extracted 3 seconds windows for all trials, because Cho2017 had a maximum trial length of 3 seconds. Note that this may not utilise the full potential of the 4-second target test data. 17 common EEG channels of the four data sets were selected (Fz, FC1, FC2, C1--C6, CP3, CP1, CPz, CP2, CP4, P1, P2, Pz). All trials are down-sampled to 200Hz. A 5th-order bandpass filter was applied between 4Hz and 32Hz, where motor imagery usually occurs. Normalisation across channels is done, followed by temporal normalisation across time steps.

Shallow ConvNet was used as the feature extractor. For both the baseline model(Shallow ConvNet) and the MF-SCSN, the feature size was aligned to 50 after the feature extractor. The common cloud network consists of three fully-connected layers with a feature size of 50 each. Before the classifiers, the separate layers consist of three fully-connected layers with a feature size of 50. All four data sets use individual classifiers according to their own tasks. During training, the target classier kept the original four classes. After prediction, feet and rest labels were combined as `other' to report the final accuracy. For all four data sets, trial numbers are balanced to 2880 trials (the size of BCIC sources). To balance training sizes, a random sampling was done for Cho2017 (originally 9880 trials). Similarly, we augment  PhysioMI (2399 trials) and target data set to 2880 trials. We used batch size $10$, learning rate  $0.001$ and weight decay factor $0.0005$ both shallow CovNet baseline and MF-SCSN. During the training of MF-SCSN, four batches of size 10 from branches (40 in total) were delivered to the cloud layers simultaneously and distributed back to each separate local branch after common feature extraction. In the target training set, subject 1-3 has 100 trials each. Subject 4-5 has 120 trials each. 20 trials each were used as the validation set for model selection. All randomization and initialization were conducted with a typical random seed for machine learning of 42 for reproducibility in the same setup and environment.

%% file: 4results.tex
\section{Results}
We tested the baseline shallow ConvNet and the MF-SCSN on the BEETL motor imagery task. Five subjects from two data sets were tested. The first three subjects (S1 S2 S3) are from the CybathlonIC data set~\cite{pmlr-v176-wei22a} collected in an online closed-loop format. The latter two subjects are from the Weibo2014 data set~\cite{weibo}. Weibo2014 uses instructions on the screen to inform subjects to perform offline motor imagery without feedback and real-time control. The CybathlonIC used the Cybathlon 2020 BCI game for data collection. The intention of controlling the virtual car in a real-world setup with real-time feedback and interference made the brain signal more complex and noisier to decode. 

\begin{figure}[h]
\centering
\vspace{ -3mm }
\includegraphics[width=0.8 \columnwidth]{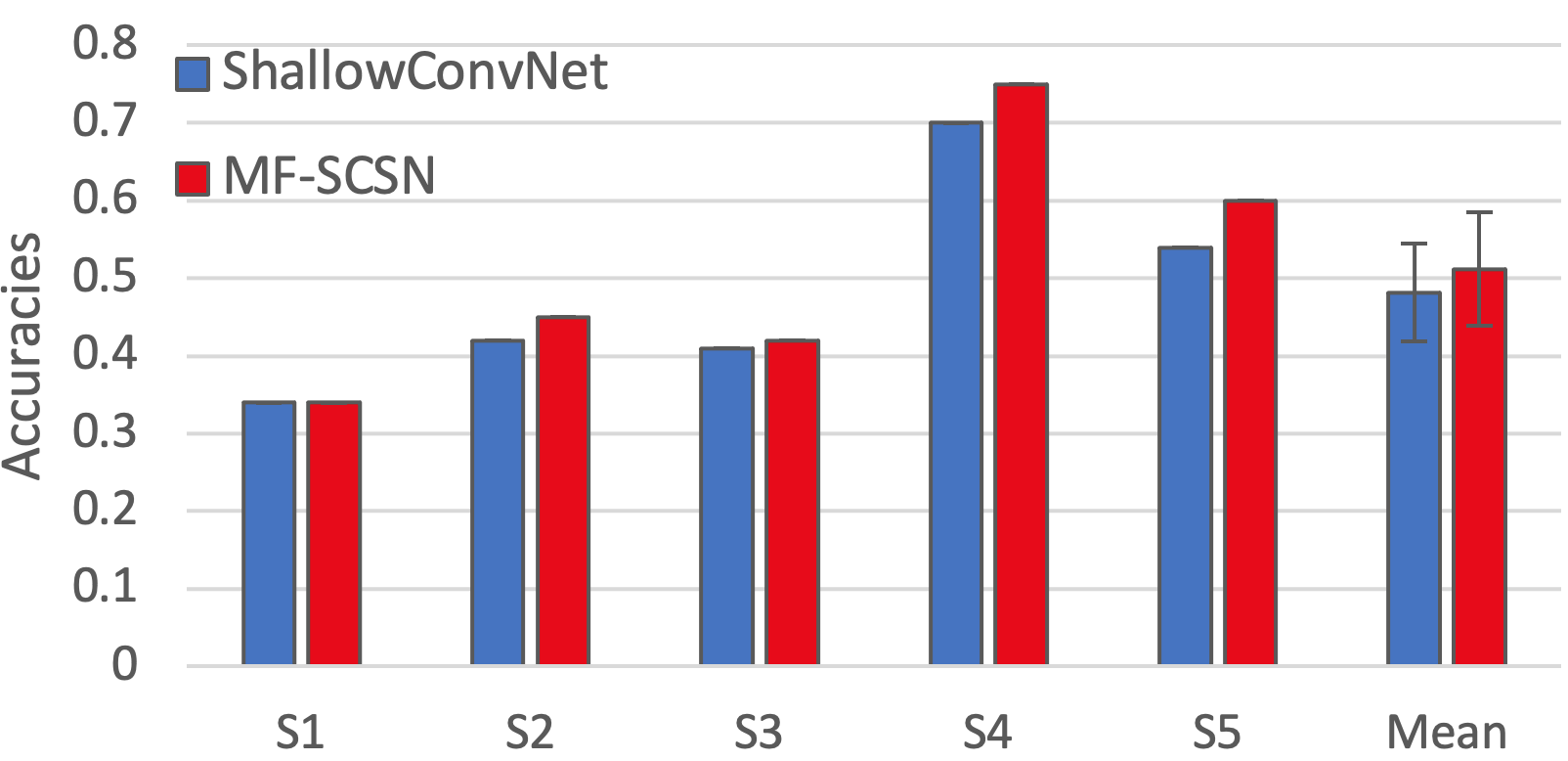} 
\vspace{-1mm}
\caption{Label-weighted decoding accuracy of subjects, their mean and standard error of the mean (SEM shown as error bar).}
\label{fig3}
\vspace{-3mm}
\end{figure}

In light of the stated differences between the above two data sets, we trained them respectively instead of pooling them together as the same target set. Prediction results were reported based on their own models. In our previous work, a significant accuracy drop could be observed by simply adding more subjects to train the ShallowConvNet~\cite{wei2021inter}. Therefore, here we used only the target subjects to train the ShallowConvNet as the baseline. For the MF-SCSN, S1-S3 are regarded as the target set and trained together with the source branches. Similarly, we trained another model combining S4 and S5. As shown in figure~\ref{fig3}, the decoding accuracies of S1-S3 are significantly lower than S4 and S5 in both the baseline method and MF-SCSN. This reflects the challenge of real-world BCI decoding and the variance of the two data sets.

The main observation of this study is that, as in figure~\ref{fig3}, the MF-SCSN outperformed the baseline methods. Among the 1000 trials (200 trials per subject), the MF-SCSN correctly classified 555 samples compared with 518 samples by the shallow ConvNet. In the 1000 testing samples, there are 250 left-hand/right-hand trials and 500 trials labelled as `others' (the feet and rest). Therefore, by giving half weight to `other' trials, weighted average decoding accuracies are computed for both methods. As shown in the last column of figure~\ref{fig3}, the baseline yielded a decoding accuracy of 48.2\%, and MF-SCSN outperformed the baseline with an accuracy of 51.2\%.

% \begin{table}[htbp]
% \centering
% \caption{Decoding result of models }
% \resizebox{8cm}{!} {
% \begin{tabular}{lllllll}
% \hline
% {\color[HTML]{000000} Network} & S1 & S2  & S3 & S4  & S5  & overall \\ \hline
% ShallowCovNet                    & 83 &  94  & 91 & 139 & 111 &  518    \\
% MF-SCSN                       & 83 & 102 & 97 & 149 & 124 & 555     \\ \hline
% \label{table1}
% \end{tabular}
% \aldo{why a table of bar charts show so much more?}
% }
% \footnotesize
% \begin{itemize}
%     \item[$\ast$] Each subject has 200 trials. The overall testing size is 1000. The best and worst performances are in bold and italics respectively.
% \end{itemize}
% \end{table}

%% file: 5discussion.tex
\section{Discussion}
Our MF-SCSN architecture outperformed the existing methods which we used as a baseline. This indicates that MF-SCSN has the potential to utilise cross-dataset federated features from different tasks to increase EEG decoding performance. We also highlight five privacy-preserving properties of the MF-SCSN.

% Testing set B increased more - 1. Set A is more complex 2. Source set data collection criteria is more similar, collected with offline instructions.
First, subjects and dataset-specific information are stored locally with data owners, which preserves data-level privacy. To preserve parameter-level privacy, feature extractors are stored locally. Another property is that local feature extractors encrypt raw data naturally, so there is no extra encryption cost. Additionally, there is no need for a protocol of decryption, because the cloud network only uses the encrypted features for transfer learning. Finally, labels and classifiers are stored and predicted locally. This also preserves inference-level privacy.

% \aldo{PLEASE no bulletpoints. generate proper text}
% \Xiaoxi{structured into a paragraph}
One limitation of this study is that some parts of the source and target data sets were discarded to align the input shape, e.g. the window length and channels. Further experiments should be conducted on transfer learning with different input shapes. A potential solution based on MF-SCSN could be exploring if the local feature extractors with different kernels could handle inputs of different shapes, by unifying output shapes of the features across branches. Another direction could be exploring the flexibility of MF-SCSN as a meta-architecture by changing the feature extractors to other models.

% The shallow ConvNet was used here, other EEG models like EEGNet~\cite{lawhern2018eegnet} or EEG-InceptionNet~\cite{santamaria2020eeg} could also be potentially plugged into the MF-SCSN.

% \aldo{please provde a proper final sentence with appropriate zoom out} 
% \Xiaoxi{summary and outlook below}
To conclude, we have designed a cross-dataset federated deep transfer learning technique which combines privacy-preserving properties and deep transfer learning. Results show that the proposed method, with the advantage of both transfer learning and privacy-preserving, outperformed the baseline CNN. Our proposed method shows the potential to utilise larger heterogeneous data sets with different tasks for transfer learning while possessing better properties of privacy-preserving across data sets and data centres.
% \aldo{nice}

% One limitation of this study is that some parts of the source and target data sets were discarded to align the input shape, e.g. the window length and channels. Further experiments should be conducted on transfer learning with different input shapes. A potential solution based on MF-SCSN could be exploring if the local feature extractors with different kernels could handle inputs of different shapes. This is possible once the output shapes of the encryption are unified across branches. Further alignment methods would be necessary to solve this more difficult transfer learning problem, e.g. euclidean alignment\cite{he2019transfer}, deep adaptation networks\cite{long2015learning} or deep set alignment\cite{zaheer2017deep}. The flexibility and robustness of MF-SCSN should also be tested as a meta-architecture, by changing the feature extractor. The shallow ConvNet was used here, other EEG models like EEGNet~\cite{lawhern2018eegnet} or EEG-InceptionNet~\cite{santamaria2020eeg} could also be potentially plugged into the MF-SCSN.